\documentclass[aps,prl,showpacs,showkeys,twocolumn]{revtex4-1}
\usepackage{graphicx,hyperref,amsmath,amsfonts,comment}
\usepackage{epstopdf,color,bm,multirow,rotating,soul}
\usepackage[T2A]{fontenc} \usepackage[utf8]{inputenc}
\usepackage[russian, english]{babel}
\begin{document}
\title[Visual media for monitoring trunk quantum networks]{
Visual media for monitoring trunk quantum networks}
\author{Aleksandr Alekseevich Litvinov$^{1,2}$,
Eugene Mihailovich Katsevman$^{2}$,
Oleg Igorevich Bannik$^{1,2}$,\\
Lenar Rishatovich Gilyazov$^{1,2}$,
Konstantin Sergeevich Melnik$^{1,2}$,
Marat Rinatovich Amirhanov$^{2}$,\\
Diana Yurevna Tarankova$^{3}$,
and
Nikolay Sergeevich Perminov$^{1,2,4,*}$}
\affiliation{$^{1}$ Kazan Quantum Center, Kazan National Research Technical University n.a. A.N.Tupolev-KAI, 10 K. Marx, Kazan 420111, Russia}
\affiliation{$^{2}$ "KAZAN QUANTUM COMMUNICATION" LLC, 10 K. Marx, Kazan 420111, Russia}
\affiliation{$^{3}$ Department of Radio-Electronics and Information-Measuring Technique, Kazan National Research Technical University n.a. A.N.Tupolev-KAI, 10 K. Marx, Kazan 420111, Russia}
\affiliation{$^{4}$ Zavoisky Physical-Technical Institute, Kazan Scientific Center of the Russian Academy of Sciences, 10/7 Sibirsky Tract, Kazan 420029, Russia}
\email{qm.kzn@ya.ru}
\begin{abstract}
In this paper, we formulate ideas for visual media for trunk quantum networks with lines over 100 km. We investigate the local trends analysis of performance parameters for experimental demonstration of long-distance quantum communication and discuss the prospects of commercialization of quantum-classical cloud security services.
\end{abstract}

\maketitle

\section{Trunk quantum link reliability}
Complex solutions \cite{vtyurina2018} in the field of quantum communications (QC) are crucial for the implementation of quantum networks capable of working in both urban and trunk standard fiber optic communication lines (FOCL). At the same time, from the point of view of communication systems, trunk quantum networks (TQN) with lines longer than 100 km and losses between nodes of more than 25 dB, where the signal-to-noise ratio cannot be considered large, are particularly hard to implementation. From the point of view of fundamental statistics, the fundamental difficulty here lies in the fact that despite the relatively low average percentage of errors, the magnitude of the error span and the variance of errors can be extremely large, which entails a low reliability of diagnosing errors when performing continuous tests of the TQN and especially large-scale TQN with a large number of nodes.

The solution to a similar problem in communication theory is the use of prognostic tools for monitoring and filtering errors, which will increase the reliability of determining the QBER (quantum bit error rate) for long-distance QC in continuous use. We also note that due to large errors for the limit passport operating regimes of QC complexes, the security of the communication complex should be determined by a special diagnostic system that is different from the diagnostic system for standard operating passport regimes. This difference in diagnostics is also due to the fact that the reliability of error determination in the limit regime effectively depends on a significantly larger number of physical factors, which are not always easy to track within the framework of a single structurally finished product. That is, a more powerful specialized diagnostic system should be supplied with the QC complex as a separate product and be able to work in the background of the TQN to track the entire history of changes in the line performance even when the QC complex in the node is turned off.

In this paper, we studied the trends and correlations of performance parameters for the first in Russia experimental long-distance QC at 143 km at 37 dB losses on standard FOCLs \cite{bannik2019noise,news2}, which exceeds 100 km (the size of the small standard line FOCLs along railways) with attenuation of more than 25 db.

\section{Noise in QC and modern statistics}
In August 2019, we experimentally demonstrated the trunk QC \cite{bannik2019noise,news2} at a distance of 143 km between the city of Kazan and the urban-type village of Apastovo in the Republic of Tatarstan using a modified quantum key distribution prototype \cite{bannik2017multinode,news1} that provides high noise immunity of lines and network nodes due to phase coding in a subcarrier wave \cite{merolla1999single}. The average secret key rate (SKR) was 12 bits per second with line losses of 37 dB at a distance of 143 km during a multi-day field test.
Earlier in \cite{stucki2009continuous}, an attempt was made to implement stable commercial long-distance QCs on a regular fiber line, where for conditions similar to our \cite{bannik2019noise,news2} conditions (37 dB loss), a ratio of key generation speed to a clock frequency of about 7 bit/s per 100 MHz was obtained, which is about 2 times lower than the values for our test 12 bit/s at 100 MHz in Kazan \cite{bannik2019noise,news2}. However, in \cite{stucki2009continuous} it was impossible to obtain a sufficiently stable operation of the QC system for several days to be sure of the reliability of the safety parameters. We assume that after solving stabilization issues (as was done, for example, in \cite{kulik2014method}), the QC complex described in \cite{stucki2009continuous,grande2018implementation} will be suitable for long-distance TQN.

Most of the new ideas \cite{sajeed2019approach} about axiomatically reliable ultrahigh quantum security are probable only within the framework of standard statistical methods and operate on essentially only one concept of only one of the subspecies of information entropy \cite{holevo1973bounds,kolesnichenko2018two}. However, it is well known that at least 20-30 basic statistical tests (NIST, U1) are used to quantify the randomness of a random number generator, which is the most significant primitive of any true quantum cryptosystem \cite{rukhin2001statistical}. Simply put, the significance level of the most traditional evidence of quantum security is no more than 1/20-1/30 ($\sim$ 5 \%) with respect to advanced statistical methods that are currently used in all high-tech areas. Therefore, the development of classification and certification ranks for QC must be carried out taking into account modern methods of statistical analysis.
In our opinion, a special place among them is occupied by the method of sequence of ranged amplitudes \cite{nigmatullin2003fluctuation}, which allows not only to increase the accuracy and speed \cite{smirnov2018sequences} of data processing for quantum systems, but also to increase the reliability of the analysis with small samples (about 100-200 points) \cite{perminov2018comparison}. The above ideas about the history of theoretical and mathematical statics in quantum informatics served as a starting point for the development of new applied methods for reliable classification and diagnostics of complex noise-like data for the basic primitives of quantum crypto-complexes \cite{perminov2018rcf,nigmatullin2019universal}.

\section*{Diagnostics of TQN}
\begin{figure}[t]
\includegraphics[width = 0.48\textwidth]{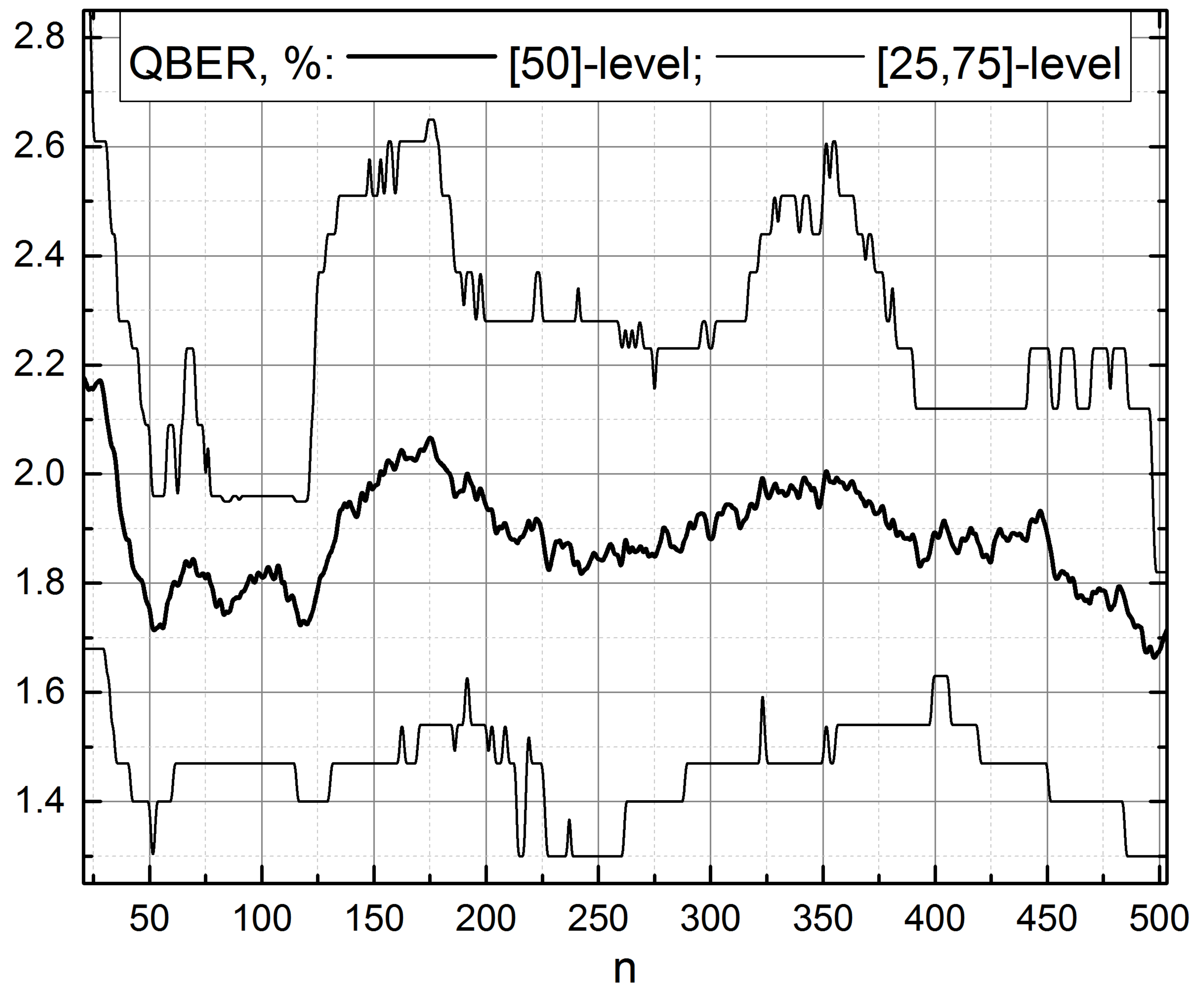}
\caption{Percentage level filters [25, 50, 75] with 50-point smoothing for QBER values, showing the presence of trends in errors during two-day test.}
\label{QBER}
\vspace{-0.5cm}
\end{figure}

\begin{figure}[t]
\includegraphics[width = 0.48\textwidth]{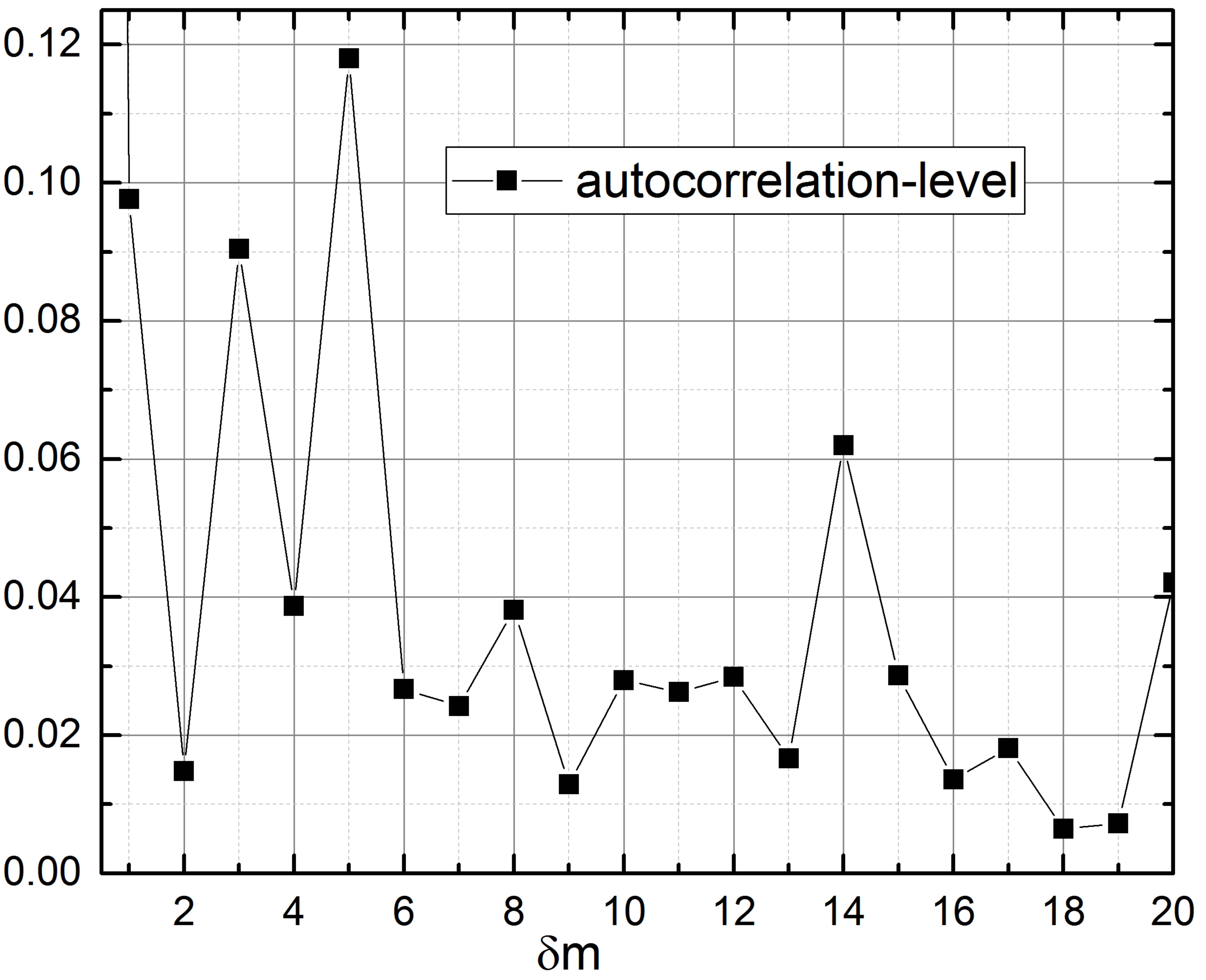}
\caption{The autocorrelation function for QBER deviations relative to the average value depending on the discrete time index $\delta m$ shows the presence of excessive correlations in the time interval [1; 5].}
\label{ACF}
\vspace{-0.5cm}
\end{figure}

During a field test of the trunk QC in Kazan \cite{bannik2019noise,news2} in August 2019, a number of QBER values were obtained (values within the acceptable QBER value were taken less than 4 \%), which even at the first examination has a trend structure that meets nonrandom factors (temperature, humidity etc). In Fig. \ref{QBER} shows percentage level filters [25, 50, 75] at 50-point smoothing for QBER values, visually showing the presence of trends in errors for two-day time series. Further detalization of nonrandomness can be further performed in a forecast format based on simple linear regressions. Accordingly, to establish the reliability of the fact of the presence of regression, you can use the autocorrelation function. In Fig. \ref{ACF} shows the autocorrelation function for QBER deviations relative to the average value depending on the discrete time index $\delta m$ (tied to a unit of measurement: about 520 points for 2 days of testing), which even more clearly shows the presence of excessive local correlations in the time interval [1; 5].
Thus, it is shown that in the main QC there are noise factors that can be predicted. This means that the specialized diagnostic system of the trunk QC should clearly predict the noise level, which in turn will increase the reliability of determining the level of interference and the safety level of the TQN.

\section*{Analysis of key times tags}
\begin{figure}[t]
\includegraphics[width = 0.48\textwidth]{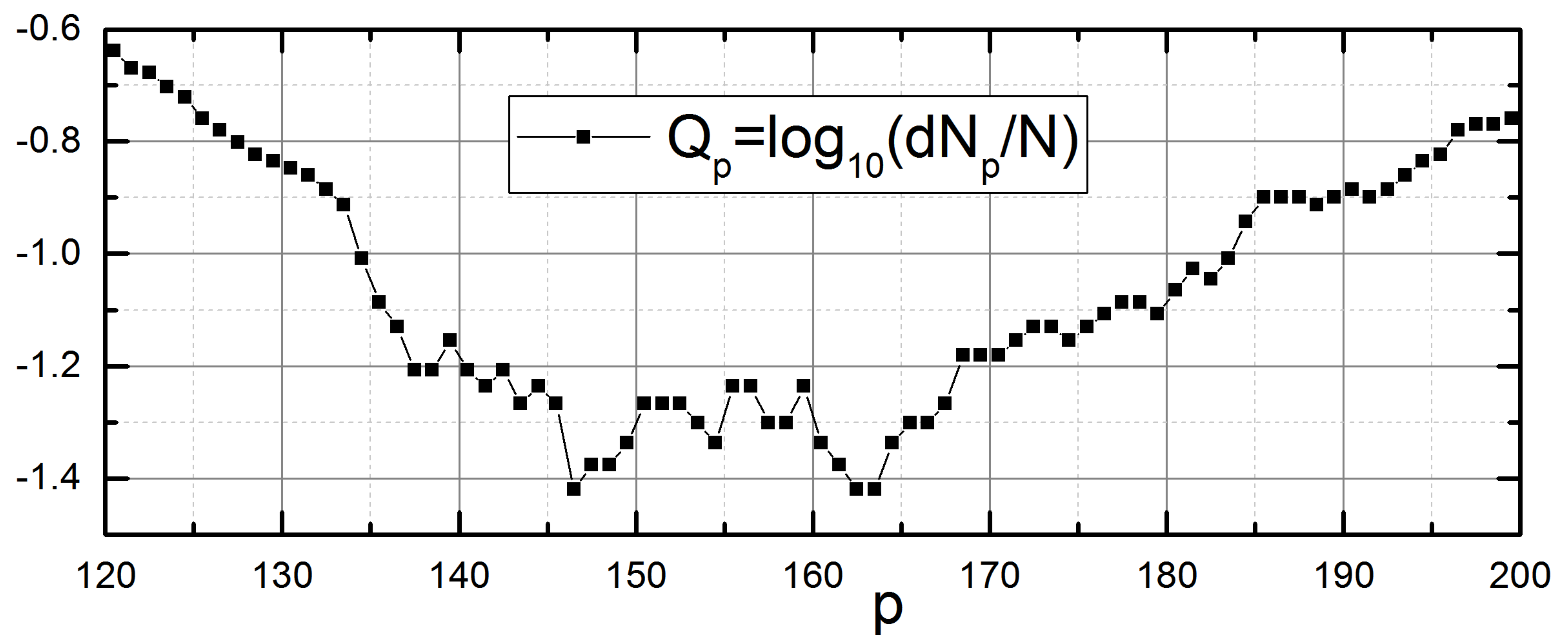}
\caption{Randomness quality factor $Q_p$ for raw key times tags for alternative measurement of the level of local correlations.}
\label{QQR}
\vspace{-0.5cm}
\end{figure}

One of the questions little studied in the context of quantum cryptography is the question of local monitoring of the difference in the time labels of bits within one raw key. Here we propose using an alternative measure of randomness for the source data for the tims tags difference $\{y_k\}$, based on the rank statistics $\{x_{k,p}=sign(y_k-p)\}$: an analog of the autocorrelation coefficient $Q_p=log_{10}( \langle ceil(x_{k,p} x_{k+1,p}) \rangle )$ with continuous index p, which gives a p-parametric family in comparison with the usual correlation coefficients \cite{perminov2019extended} (here $ceil(x)$ is the integer part of a $x$). In Fig. \ref{QQR} shows the quality factor of randomness $Q_p$ for raw key times tags for alternatively measuring the level of local correlations.
Note that the decimal logarithm of the standard Pearson autocorrelation coefficient gives a value of -1.06, significantly different compared to $min_p(Q_p)=-1.42$, which shows the additional capabilities of the new technique.

As predictive regression is built on the redundancy of Pearson autocorrelation, it is also possible to construct an alternative predictive regression for an alternative measure of correlations. At the same time, differences in the Pearson correlation levels and the introduced correlation $Q_p$ actually quantitatively show which prognostic model will work better in this particular situation. Thus, advanced rank statistics when monitoring the quantum randomness of keys in trunk QCs open up new possibilities for predicting noise and failures in the TQN.

\section*{Conclusion}
We considered the question of the fundamental applicability of traditional statistical methods for monitoring long-range quantum communication systems operating in extreme passport regimes under high noise conditions. Field tests showed the inapplicability of standard criteria for quantum informatics for a deep detailed applied assessment of the reliability of noise detection. Nevertheless, for trunk QCs, there is a need to introduce some basic classification and certification safety criteria \cite{grande2018implementation,drahi2019certified}.
It is worth noting the importance of prognostic monitoring systems in the context of building a large-scale national quantum network, what is based on the first continuous test of main QC in Russia \cite{bannik2019noise,news2}. Moreover, the diagnosis of errors in the extreme passport regime depends on a large number of physical factors that are difficult to track within the framework of only one structurally finished product.

Therefore, we need an expanded system of visual diagnostics, which should be supplied with the QC complex and be able to work in the background of the TQN to track the entire history of changes in the line performance even when the QC complex in the node is disabled. Such a quantum-classical diagnostic subnet in the TQN, capable of diagnosing even small noises in the network in the absence of a priori information about the type of interference, opens up prospects for the commercialization of quantum-classical cloud services for robust information protection.

\section*{Trunk QC}
Noise immunity and active intelligent stabilization \cite{kulik2014method,kulik2014line,balygin2016active} are very important characteristics of systems for ultra-long distance QCs.
The characteristics allow to classify the existing long-distance QCs \cite{molotkov2008resistance,molotkov2011solution} by using extensive  quantum analysis \cite{molotkov2016complexity} (traditional for cryptography) and intelligent noise-analysis \cite{nigmatullin2017general,nigmatullin2019universal,nigmatullin2019detection,nigmatullin2016forecasting} (traditional for steganography, friend identification systems and general type complex systems).
For solution of the problems of noise immunity in relation to the influences on both the  fiber communication line and QKD apparatus in the network nodes, we must accordingly abandon the use of interferometers \cite{molotkov2004multiplex} reducing vibration resistance of nodes and polarization coding sensitive to deformations, vibration and temperature changes in the fiber.

The unique noise immunity is inherent to the subcarrier wave phase-coding protocol \cite{merolla1999single} what distinguishes this approach from many others \cite{wehner2018quantum,djordjevic2019recent,razavi2019quantum}.
A quantum signal is not directly generated by the source in the one-way QKD system of subcarrier frequencies, but is created as a result of phase modulation of carrier wave at Alice and Bob sides \cite{merolla1999single}. 
A description of a typical technological implementation of this protocol can be found in \cite{gleim2016secure}. In \cite{bannik2019noise,news2} we offered a modified variant of this QKD-system.

For the field tests in \cite{bannik2019noise,news2}, we used Kazan-Apastovo regular fiber link with a length of 143 km and losses of 37 dB in the quantum channel and 45 dB in the synchronization channel with erbium-doped fiber amplifier. The total time for continuous testing was 16.5 hours. 
About 700 kbit of key information was generated in the quantum channel with an average value of the secret quantum key generation rate of 12 bps.
The value of QBER was in the stable range from 0.5 $\%$ to 3.5 $\%$, on average, the value of QBER $\sim$ 2 $\%$ (see Fig. \ref{QBER}). 
This generation rate allows to change the 256-bit encryption key up to two times per minute. The data of the stable experimental demonstration show the possible using of the proposed QKD system in commercial long-range QCs.

Basic characteristics of the tests of our QKD system for urban quantum networks \cite{bannik2017multinode,news1} and intercity trunk lines \cite{bannik2019noise,news2} are presented in Tab. \ref{T:param}. For comparison, Tab. \ref{T:param} also presents the basic characteristics of the well-known long-range QKD systems discussed in the introduction which were also tested under real conditions \cite{notes1}. 

\begin{table}[t]
\begin{tabular}{|c|c|c|c|c|}
\hline
\multicolumn{5}{|c|}{Kazan quantum lines} \\ \hline
Length& SKR & Detector & Loss (dB); & Group \\
(km) & (bps) &  & QBER (\%) &  \\ \hline
12 & $2 \cdot10^4$ & $^*$SPAD & 7; 4 & Russia \\ \hline
143 & 12 & $^\dagger$SSPD & 37; 2 & Russia \\ \hline
\multicolumn{5}{|c|}{International commercial systems} \\ \hline
67 & 60 & \cite{stucki2002quantum} SPAD & 14; 6 & Switzerland \\ \hline
45 & $3 \cdot10^5$ & \cite{dixon2015high} SPAD & 14; 4 & Japan \\ \hline
66 & $5 \cdot10^5$ & \cite{mao2018integrating} SPAD & 21; 5 & China \\ \hline
97 & 800 & \cite{tanaka2008ultra} SSPD & 33; 3 & Japan \\ \hline
90 & $1 \cdot10^3$ & \cite{shimizu2013performance} SSPD & 30; 3 & Japan \\ \hline
\end{tabular}
\caption{Basic characteristics of the proposed QKD system: $^*$Kazan city quantum network (Kazan, May 2017 \cite{bannik2017multinode,news1}), $^\dagger$intercity long-distance QC (Kazan-Apastovo, August 2019 \cite{bannik2019noise,news2}).}
\label{T:param}
\vspace{0cm}
\end{table}
\noindent
It is worth noting that in \cite{bacco2019field,choi2014field} and \cite{walenta2015towards,boaron2018secure} tests of various QKD systems at shorter distances and using ultra-low loss fibers are presented.

\section{Innovation outlook in trunk QC}
Based on the demonstrated in \cite{bannik2019noise,bannik2017multinode,news1,news2} QKD prototype, we propose the concept of a universal QC complex $"$Kazan-Q1$"$ (see outlook Tab. \ref{T:complex} and patents \cite{patent1,patent2,patent3,patent4,patent5}) which combines the technological simple solutions, noise immunity of the subcarrier wave coding and robust quantum-classical information protection based on the extended statistical data mining against new types of quantum attacks \cite{bykovsky2018quantum}.

\begin{table}[t]
\begin{tabular}{|c|c||c|c|}
\hline
Name & Kazan-Q1 & Detector & SSPD \\
 &  &  &  or SPAD \\ \hline
Coding & Phase in & Active & Phase \\
 & subcarrier wave & element & modulator \\ \hline
Security & Quantum and & QRNG & On PD with \\
 & classical & & robust-defense \\ \hline
Noise & Network nodes & Decoys and & Amplitude \\
immunity & and lines & diagnostics & modulator and AI \\ \hline 
\end{tabular}
\caption{Outlook for the complex $"$Kazan-Q1$"$. Here: QRNG: quantum random number generator; PD: PIN-photodiode; AI: artificial intelligence.}
\label{T:complex}
\vspace{-0.3cm}
\end{table}
\noindent
We see the following distinctive features and prospects for the development of the long-distance QC complex $"$Kazan-Q1$"$ in the context of world achievements in the area of innovative quantum technologies:

1) statistical monitoring with AI for keys and electronic components based on the robust nonparametric criteria \cite{nigmatullin2010new,perminov2018rcf,nigmatullin2019universal,smirnov2018sequences,perminov2017sra};

2) compact \cite{melnik2018using,zhang2019integrated,eriksson2019wavelength} planar implementation \cite{orieux2016recent} without using interferometers for the main components \cite{acin2018quantum}:
beam splitter  \cite{desiatov2019ultra,tao2008cascade,thomaschewski2019plasmonic}, planar phase modulator \cite{ren2019integrated}, high purity QRNG on PD \cite{campbell2016recent} with robust-defense \cite{perminov2018rcf,avesani2018source,ioannou2018much,grangier2018quantum,drahi2019certified};

3) implementation of advanced decoy states \cite{molotkov2019there,huang2018quantum} in a complex using intelligent monitoring methods \cite{perminov2018rcf,nigmatullin2019universal} for quantum diagnostics of network integrity and intrusion detection
\cite{kravtsov2018relativistic,gaidash2019countermeasures,gaidash2019methods};

4) advanced post-processing \cite{jia2019quantum} for error correction and modeling in optical simulator for testing and specialization of software for specific urban conditions.

Note that at the moment, the topic of data series monitoring in quantum communications raised in this work is still purely debatable and more characteristic of such a field as, for example, econometric technical analysis of time series.

\section*{Acknowledgments}
The team of authors expresses special gratitude to the professor of the Department of Radio-Electronics and Information-Measuring Technique KNRTU-KAI Raoul Rashidovich Nigmatullin for discussing the topic of nonparametric rank criteria in physics. Research of noise in the area of photonics and quantum technologies is financially supported by a grant of the Government of the Russian Federation, project No. 14.Z50.31.0040, February 17, 2017 (experiment and nonparametric analysis – AAL, OIB, LRG, KSM, NSP). The work is also partially supported in the framework of the budget theme of the laboratory of Quantum Optics and Informatics of Zavoisky Physical-Technical Institute (numerical modeling in quantum informatics – NSP, DYT).

\newpage

\section{Information about authors}
\noindent
\textbf{Aleksandr Alekseevich Litvinov},\\
(b. 1985), in 2008 graduated the department of Theory of Relativity and Gravity of the Physics Department of Kazan Federal University in the direction of “Physics”, engineer at the Kazan Quantum Center of the KNRTU-KAI.\\
\textit{Area of interest:} quantum communications, optoelectronics, robust methods, quantum memory, software, economics.\\
\textit{E-mail:} litvinov85@gmail.com\\
\\
\noindent
\textbf{Eugene Mihailovich Katsevman},\\
(b. 1984), In 2009 he graduated from the Faculty of Computational Mathematics and Cybernetics of Kazan Federal University with a degree in Mathematics, systems programmer in a private company.\\
\textit{Area of interest:} machine learning, network technologies, system design and automation.\\
\textit{E-mail:} eugene.katsevman@gmail.com\\
\\
\noindent
\textbf{Oleg Igorevich Bannik},\\
(b. 1988), in 2012 graduated from the faculty of electronics of Saint Petersburg Electrotechnical University in the direction of "Electronics and Microelectronics", researcher at the Kazan Quantum Center of the KNRTU-KAI.\\
\textit{Area of interest:} quantum communications, optoelectronics, photonics.\\
\textit{E-mail:} olegbannik@gmail.com\\
\\
\noindent
\textbf{Lenar Rishatovich Gilyazov},\\
(b. 1985), in 2008 graduated from the the Physics Department of Kazan Federal University, researcher at the Kazan Quantum Center of the KNRTU-KAI.\\
\textit{Area of interest:} quantum communications, optoelectronics, photonics.\\
\textit{E-mail:} lgilyazo@mail.ru\\
\\
\noindent
\textbf{Konstantin Sergeevich Melnik},\\
(b. 1993), in 2018 graduated from Institute of Radio Electronics and Telecommunications of Kazan National Research Technical University in the direction of "Radio Engineering", engineer at the Kazan Quantum Center of the KNRTU-KAI.\\
\textit{Area of interest:} quantum communications, optoelectronics, photonics.\\
\textit{E-mail:} mkostyk93@mail.ru\\
\\
\noindent
\textbf{Marat Rinatovich Amirhanov},\\
(b. 1985), in 2008 he graduated from the faculty of power engineering Kazan National Research Technological University.\\
\textit{Area of interest:} quantum communications, network technologies, automation and programming.\\
\textit{E-mail:} m.amirhanov85@gmail.com\\
\\
\noindent
\textbf{Diana Yurevna Tarankova},\\
(b. 1995), since 2017 a student at the Institute of Radio Electronics and Telecommunications of Kazan National Research Technical University.\\
\textit{Area of interest:} programming, photonics, network technologies.\\
\textit{E-mail:} tarankovadyu@ya.ru\\
\\
\noindent
\textbf{Nikolay Sergeevich Perminov},\\
(b. 1985), in 2008 graduated the department of Theory of Relativity and Gravity of the Physics Department of Kazan Federal University in the direction of “Physics”, researcher at the Kazan Quantum Center of the KNRTU-KAI.\\
\textit{Area of interest:} optimal control, quantum informatics, statistics, software, economics.\\
\textit{E-mail:} qm.kzn@ya.ru\\

\vspace{-1.0cm}

\bibliographystyle{ieeetr}
\bibliography{VS_QN}

\end{document}